\documentclass[aps,prl,twocolumn,superscriptaddress,showpacs]{revtex4} 

\usepackage{graphicx}  
\usepackage{dcolumn}   
\usepackage{bm}        
\usepackage{amssymb}   

\newcommand{\met}{\,/\!\!\!\!E_T}
\newcommand{\me}{\,/\!\!\!\!E}
\newcommand{\mvis}{m_{vis}}

\begin{document}

\title{Search for neutral MSSM Higgs bosons decaying to $\tau$ pairs in $p\bar{p}$
collisions at $\sqrt{s}=1.96$~TeV}

\affiliation{Institute of Physics, Academia Sinica, Taipei, Taiwan 11529, Republic of China}
\affiliation{Argonne National Laboratory, Argonne, Illinois 60439}
\affiliation{Institut de Fisica d'Altes Energies, Universitat Autonoma de Barcelona, E-08193, Bellaterra (Barcelona), Spain}
\affiliation{Baylor University, Waco, Texas  76798}
\affiliation{Istituto Nazionale di Fisica Nucleare, University of Bologna, I-40127 Bologna, Italy}
\affiliation{Brandeis University, Waltham, Massachusetts 02254}
\affiliation{University of California, Davis, Davis, California  95616}
\affiliation{University of California, Los Angeles, Los Angeles, California  90024}
\affiliation{University of California, San Diego, La Jolla, California  92093}
\affiliation{University of California, Santa Barbara, Santa Barbara, California 93106}
\affiliation{Instituto de Fisica de Cantabria, CSIC-University of Cantabria, 39005 Santander, Spain}
\affiliation{Carnegie Mellon University, Pittsburgh, PA  15213}
\affiliation{Enrico Fermi Institute, University of Chicago, Chicago, Illinois 60637}
\affiliation{Joint Institute for Nuclear Research, RU-141980 Dubna, Russia}
\affiliation{Duke University, Durham, North Carolina  27708}
\affiliation{Fermi National Accelerator Laboratory, Batavia, Illinois 60510}
\affiliation{University of Florida, Gainesville, Florida  32611}
\affiliation{Laboratori Nazionali di Frascati, Istituto Nazionale di Fisica Nucleare, I-00044 Frascati, Italy}
\affiliation{University of Geneva, CH-1211 Geneva 4, Switzerland}
\affiliation{Glasgow University, Glasgow G12 8QQ, United Kingdom}
\affiliation{Harvard University, Cambridge, Massachusetts 02138}
\affiliation{Division of High Energy Physics, Department of Physics, University of Helsinki and Helsinki Institute of Physics, FIN-00014, Helsinki, Finland}
\affiliation{University of Illinois, Urbana, Illinois 61801}
\affiliation{The Johns Hopkins University, Baltimore, Maryland 21218}
\affiliation{Institut f\"{u}r Experimentelle Kernphysik, Universit\"{a}t Karlsruhe, 76128 Karlsruhe, Germany}
\affiliation{High Energy Accelerator Research Organization (KEK), Tsukuba, Ibaraki 305, Japan}
\affiliation{Center for High Energy Physics: Kyungpook National University, Taegu 702-701; Seoul National University, Seoul 151-742; and SungKyunKwan University, Suwon 440-746; Korea}
\affiliation{Ernest Orlando Lawrence Berkeley National Laboratory, Berkeley, California 94720}
\affiliation{University of Liverpool, Liverpool L69 7ZE, United Kingdom}
\affiliation{University College London, London WC1E 6BT, United Kingdom}
\affiliation{Massachusetts Institute of Technology, Cambridge, Massachusetts  02139}
\affiliation{Institute of Particle Physics: McGill University, Montr\'{e}al, Canada H3A~2T8; and University of Toronto, Toronto, Canada M5S~1A7}
\affiliation{University of Michigan, Ann Arbor, Michigan 48109}
\affiliation{Michigan State University, East Lansing, Michigan  48824}
\affiliation{Institution for Theoretical and Experimental Physics, ITEP, Moscow 117259, Russia}
\affiliation{University of New Mexico, Albuquerque, New Mexico 87131}
\affiliation{Northwestern University, Evanston, Illinois  60208}
\affiliation{The Ohio State University, Columbus, Ohio  43210}
\affiliation{Okayama University, Okayama 700-8530, Japan}
\affiliation{Osaka City University, Osaka 588, Japan}
\affiliation{University of Oxford, Oxford OX1 3RH, United Kingdom}
\affiliation{University of Padova, Istituto Nazionale di Fisica Nucleare, Sezione di Padova-Trento, I-35131 Padova, Italy}
\affiliation{University of Pennsylvania, Philadelphia, Pennsylvania 19104}
\affiliation{Istituto Nazionale di Fisica Nucleare Pisa, Universities of Pisa, Siena and Scuola Normale Superiore, I-56127 Pisa, Italy}
\affiliation{University of Pittsburgh, Pittsburgh, Pennsylvania 15260}
\affiliation{Purdue University, West Lafayette, Indiana 47907}
\affiliation{University of Rochester, Rochester, New York 14627}
\affiliation{The Rockefeller University, New York, New York 10021}
\affiliation{Istituto Nazionale di Fisica Nucleare, Sezione di Roma 1, University of Rome ``La Sapienza," I-00185 Roma, Italy}
\affiliation{Rutgers University, Piscataway, New Jersey 08855}
\affiliation{Texas A\&M University, College Station, Texas 77843}
\affiliation{Istituto Nazionale di Fisica Nucleare, University of Trieste/\ Udine, Italy}
\affiliation{University of Tsukuba, Tsukuba, Ibaraki 305, Japan}
\affiliation{Tufts University, Medford, Massachusetts 02155}
\affiliation{Waseda University, Tokyo 169, Japan}
\affiliation{Wayne State University, Detroit, Michigan  48201}
\affiliation{University of Wisconsin, Madison, Wisconsin 53706}
\affiliation{Yale University, New Haven, Connecticut 06520}
\author{A.~Abulencia}
\affiliation{University of Illinois, Urbana, Illinois 61801}
\author{D.~Acosta}
\affiliation{University of Florida, Gainesville, Florida  32611}
\author{J.~Adelman}
\affiliation{Enrico Fermi Institute, University of Chicago, Chicago, Illinois 60637}
\author{T.~Affolder}
\affiliation{University of California, Santa Barbara, Santa Barbara, California 93106}
\author{T.~Akimoto}
\affiliation{University of Tsukuba, Tsukuba, Ibaraki 305, Japan}
\author{M.G.~Albrow}
\affiliation{Fermi National Accelerator Laboratory, Batavia, Illinois 60510}
\author{D.~Ambrose}
\affiliation{Fermi National Accelerator Laboratory, Batavia, Illinois 60510}
\author{S.~Amerio}
\affiliation{University of Padova, Istituto Nazionale di Fisica Nucleare, Sezione di Padova-Trento, I-35131 Padova, Italy}
\author{D.~Amidei}
\affiliation{University of Michigan, Ann Arbor, Michigan 48109}
\author{A.~Anastassov}
\affiliation{Rutgers University, Piscataway, New Jersey 08855}
% ...
\collaboration{CDF Collaboration}
\noaffiliation

\date{\today}

\begin{abstract}
We present a search for neutral supersymmetric Higgs bosons
decaying to $\tau$ pairs produced in $p\bar{p}$ collisions at
$\sqrt s$ = 1.96~TeV. The data, corresponding to 310~pb$^{-1}$
integrated luminosity, were collected with the CDF~II detector at
the Tevatron collider at Fermilab. No significant excess above the
standard model backgrounds is observed. We set exclusion limits on
the production cross section times branching fraction to $\tau$
pairs for Higgs masses in the range from 90 to 250~GeV/$c^2$.
\end{abstract}


\maketitle

One of the outstanding questions in particle physics is the
dynamics of electroweak (EW) symmetry breaking and the origin of
particle masses. In the standard model (SM), EW symmetry is
spontaneously broken through the Higgs mechanism~\cite{higgs},
which predicts the existence of a massive scalar Higgs boson $h_{SM}$.
Theoretical difficulties
related to divergent radiative corrections to the $h_{SM}$ mass
have natural solutions in supersymmetric (SUSY)
models~\cite{martin}.

The minimal supersymmetric extension of the standard model
(MSSM)~\cite{mssm} is the simplest realistic SUSY theory.
The Higgs sector in the MSSM consists of two charged and three
neutral scalar bosons. Assuming $CP$-invariance, one of the
neutral bosons ($A$) is $CP$-odd, and the other two ($h$,$H$) are
$CP$-even. Throughout this Letter, we use $h$ ($H$) for the lighter
(heavier) $CP$-even neutral Higgs boson, and $\phi$ to denote any
of $h$,$H$,$A$. At tree level, the MSSM Higgs bosons are described
by the mass of $A$ ($m_A$), and
$\tan\beta=v_u/v_d$, where $v_u$, $v_d$ are the vacuum expectation
values of the neutral Higgs fields that couple to up-type and
down-type fermions, respectively. The Yukawa couplings of $A$ to
down-type fermions (such as the {\sl b} quark and $\tau$) are
enhanced by a factor of $\tan\beta$ relative to the SM. For large
$\tan\beta$ one of the $CP$-even bosons is nearly mass-degenerate
with $A$ and has similar couplings.
The dominant production mechanisms of neutral MSSM Higgs
bosons at hadron colliders are gluon fusion~\cite{dawson} and
$b\bar{b}$ fusion~\cite{maltoni,harlander}. The leading decay
modes of $A$ and the corresponding mass-degenerate $CP$-even Higgs
boson are $\phi\to b\bar{b}$ ($\sim90$\%) and
$\phi\to\tau\tau$ ($\sim10$\%).

In this Letter we present the results of a search for neutral MSSM
Higgs bosons produced in $p\bar{p}$ collisions at
$\sqrt{s}$=1.96~TeV. The data sample was collected with the CDF II
detector at the Fermilab Tevatron between 2002 and 2004 and has an
integrated luminosity of 310~pb$^{-1}$. The search is performed in
the $\phi\to\tau\tau$ decay channel for $90 < m_A <
250$~GeV/$c^2$.
One $\tau$ is detected in the decay to an $e$ or $\mu$ and
neutrinos, and the other in the decay to hadrons and a neutrino.
In the following, we use $\tau_e$, $\tau_{\mu}$, and $\tau_{had}$
as shorthand notations for the decay modes $\tau \to
e\nu_e\nu_{\tau}$, $\tau \to \mu\nu_{\mu}\nu_{\tau}$, and $\tau
\to hadrons~\nu_{\tau}$, respectively. Previous and related
searches in the di-$\tau$ channel are presented in
\cite{run1,hmtt}.

CDF~II~\cite{CDFdetector} is a general purpose detector with
tracking and calorimetry. The tracking system consists of silicon
micro-strip detectors and a cylindrical wire drift chamber. It is
immersed in a 1.4~T magnetic field produced by a superconducting
solenoid. Electromagnetic (EM) and hadronic (HAD) sampling
calorimeters are located outside the solenoid and cover detector
pseudorapidity $|\eta|<3.6$, where $\eta=-\ln(\tan(\theta /2))$
and $\theta$ is the polar angle with respect to the proton beam.
The calorimeters are divided into towers with projective geometry.
A central electromagnetic shower maximum detector (CES) consisting
of proportional chambers with anode wires parallel to the beam
axis and orthogonal cathode strips is embedded in the EM
calorimeter at a depth of six radiation lengths. The CES is used
to determine the position of EM showers with spatial resolution of
$\sim0.5$~cm. Muons are identified by a system of drift chambers
located outside the calorimeter volume with combined coverage
extending to $|\eta|<1.5$. The luminosity is measured by gas
Cherenkov counters located in the detector forward and backward
regions ($3.7<|\eta|<4.7$) with 6\% precision~\cite{luminosity}.

The search for $\phi\to\tau\tau$ requires detection of an $e$ or $\mu$
(from $\tau_e$, $\tau_{\mu}$) and the reconstruction of the
$\tau_{had}$ decay products. Events are pre-selected with ``lepton
plus track" triggers~\cite{tauTrig}. The triggers require a lepton
($e,\mu$) candidate and another track, both pointing to the
central calorimeter ($|\eta|\lesssim 1.0$) and having azimuthal
separation $\Delta\varphi>10^\circ$.
The overall trigger efficiency for signal events passing the selection
criteria described below is greater than 90\%.
The algorithms for $e$ and $\mu$ identification are described in
detail in~\cite{CDFdetector}.
The vector sum of the transverse momenta~\cite{pt} of the neutrinos from $\tau$
decays appears as missing transverse energy ($\met$), determined
from the imbalance of energy deposition in the
calorimeter~\cite{metDef}.
The decay products in $\tau_{had}$ form narrow jets with
low multiplicity of neutral and charged particles. The positions
and energies of $\pi^0$'s and photons are reconstructed with the
CES detector and EM calorimeter, respectively.
In this search we do not distinguish reconstructed photons and $\pi^0$'s,
and all neutrals are assumed to be $\pi^0$'s.
We use a variable-size signal cone and an isolation annulus~\cite{hmtt}
to identify $\tau_{had}$ candidates.
The four-momentum of $\tau_{had}$ is calculated from
tracks and $\pi^0$'s contained in the signal cone.
Particles in the isolation annulus are used to impose requirements that
discriminate against quark and gluon jets:
the scalar sum of the $p_T$ of tracks (sum of $E_T$ of $\pi^0$'s)
in the isolation annulus is required to be less than
1~GeV/$c$ (1~GeV). We select $\tau_{had}$ candidates with one or
three tracks in the signal cone ($N^{trk}_{sig}=1,3$) with
$p_T>1$~GeV/$c$ , consistent with the dominant $\tau$ decay modes.
In the $N^{trk}_{sig}=3$ case the sum of the electric charges
must be equal to $\pm1$. The invariant mass of the hadronic system
is required to be less than 1.8~GeV/$c^2$.
Electrons are rejected by imposing the condition
$(E^{cl}/P^{trk}_{sig})(0.95 - f)>0.1$, where $E^{cl}$ is the energy
in the calorimeter cluster produced by the $\tau_{had}$ candidate,
$f$ is the ratio of electromagnetic to hadronic energy in the cluster, and
$P^{trk}_{sig}$ is the scalar sum of track momenta in the signal
cone. Muons are suppressed by requiring $E^{cl}_T>15$~GeV. The
$\tau_{had}$ identification efficiency increases from 38\% at
transverse momentum of the hadronic system
$p_T^{had}=15$~GeV/$c$ to $\sim46\%$ for
$p_T^{had}\gtrsim25$~GeV/$c$. The probability for misidentifying a
quark or gluon jet as $\tau_{had}$ is measured using jet data
samples. It is $\sim1.5\%$ for jet transverse energy
$E_T^{jet}$=20~GeV, dropping to $\sim0.1\%$ for $E_T^{jet}$=100~GeV.

The acceptances for signal and most of the backgrounds are
determined from samples of Monte Carlo
(MC) simulated events produced by the PYTHIA event
generator~\cite{pythia} with CTEQ5L~\cite{cteq5} parton
distribution functions (PDF's). Tau decays are simulated by the
TAUOLA package~\cite{tauola}. Detector response is simulated with
a GEANT-based~\cite{geant} model of the detector.

The dominant (and irreducible) background in the final sample of
selected events is from inclusive $Z/\gamma^*$ production
with subsequent decays to $\tau$ pairs. It is estimated using MC
simulated events with normalization corresponding to
$\sigma(p\bar{p}\to Z/\gamma^*) \times BR(Z/\gamma^* \to ll) =
254.9$~pb in the di-lepton mass region
$66<m_{ll}<116$~GeV/$c^2$~\cite{zCrossSection}.
The second largest background contribution comes from processes
with quark or gluon jets misidentified as $\tau_{had}$, such as
di-jet and multi-jet, $W$+jets and $\gamma$+jets production. These
backgrounds are estimated from the data by applying jet$\to
\tau_{had}$ misidentification rates to jets in events that pass
all selection criteria except for $\tau_{had}$ identification.
The validity of the predictions is verified using independent data
samples representing the background processes.
The third group of backgrounds includes $Z/\gamma^*\to ll$
($l=e,\mu$), $WW$, $WZ$, $ZZ$, and $t\bar{t}$ production. Their
contributions are determined from MC samples normalized to the
theoretical cross sections.

The events in the $\tau_e\tau_{had}$ ($\tau_{\mu}\tau_{had}$)
channel are selected by requiring one $e$ ($\mu$) candidate with
$p^{e(\mu)}_T>10$~GeV/$c$, and
one $\tau_{had}$ candidate with $p^{had}_T>15$~GeV/$c$ and
opposite electric charge.
Low-energy multi-jet backgrounds are suppressed by rejecting events with
$|p^{e(\mu)}_T|+|p^{had}_T|+|\met|<50$~GeV. Backgrounds from $W$ +
jet events are suppressed by imposing a requirement on the relative
directions of the visible $\tau$ decay products and $\met$.
We define a unit vector $\hat{\zeta}$ along the bisector
of the angle between the directions of $e$ ($\mu$) and
$\tau_{had}$ in the transverse plane.
The projections $p_{\zeta}^{vis} =
(\overrightarrow{p}_{e(\mu)} +
\overrightarrow{p}_{had})\cdot{\hat{\zeta}}$ and $p_{\zeta}^{\met}
=  \overrightarrow{\me}_T \cdot\hat{\zeta}$ are required to
satisfy $p_{\zeta}^{\met}>0.6 p_{\zeta}^{vis} - 10$~GeV/$c$. This
condition removes $\sim 85$\% of the $W$ + jet events passing the
other selection criteria, while retaining $\sim 95$\% of the signal.
To suppress backgrounds from $Z\to ll$ decays with a misidentified lepton,
we do not accept events  with invariant mass of an $e$ ($\mu$) and a single-track
$\tau_{had}$ candidate within 10~GeV/$c^2$ of the $Z$ mass.
The combined signal acceptance for a Higgs boson of mass
90~GeV/$c^2$ (250~GeV/$c^2$) in the $\tau_e\tau_{had}$ and
$\tau_{\mu} \tau_{had}$ channels is 0.8\% (2.0\%).

The systematic uncertainties for particle identification
efficiency are 3.5\% ($\tau_{had}$), 1.3\%  ($e$), and 4.6\%
($\mu$). The uncertainties in trigger efficiency for
the $\tau_e\tau_{had}$ and $\tau_{\mu}\tau_{had}$ channels are
2.1\% and 1.4\%, respectively.
The uncertainty in the determination of backgrounds due
to jet$\to\tau$  misidentification is 20\%, resulting in 3\%
effect on the total background estimate. The systematic
uncertainty in signal acceptance from event-level cuts is less
than 2\%. The imprecise knowledge of the PDF's introduces an
additional 5.7\% uncertainty on signal
acceptance~\cite{pdfUncert}.

\begin{figure}
\includegraphics[width=0.95\columnwidth]{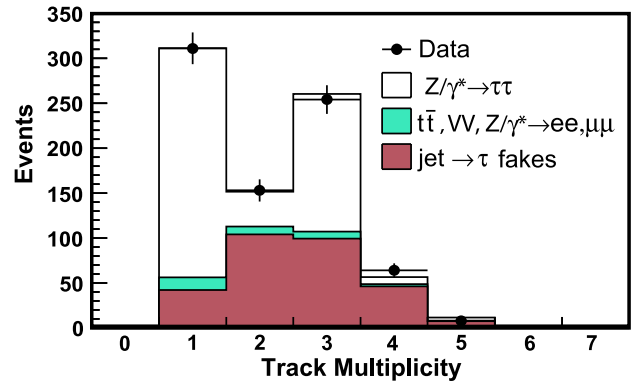}
\caption{Track multiplicity for hadronically decaying tau
candidates before applying the opposite charge and $N_{trk}^{sig}$
requirements.} \label{fig:nTracks}
\end{figure}

Figure~\ref{fig:nTracks} shows the track multiplicity distribution
for $\tau_{had}$ candidates in the data, along
with the background predictions.
The characteristic enhancement in the one- and three-track bins
clearly shows the contribution from events with $\tau_{had}$ in
the final state.
The total number of expected events from SM processes after
applying all selection criteria is $N_{SM}=496 \pm 5(stat) \pm
28(sys) \pm 25(lumi)$.
The contributions from $Z/\gamma^* \to \tau\tau$, backgrounds with
jet$\to \tau$ misidentification, and all remaining background
sources are 405, 75, and 16, respectively.

\begin{figure}
\includegraphics[width=0.95\columnwidth]{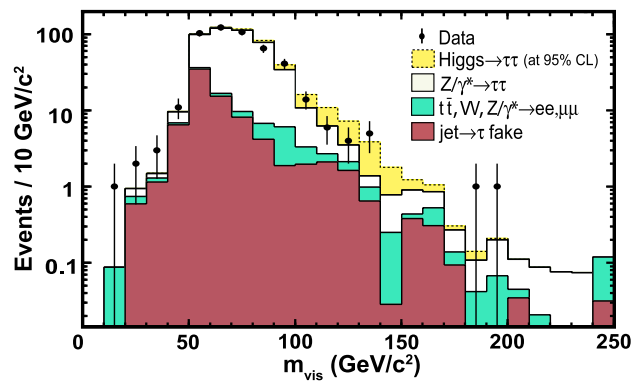}
\caption{Example fit of the $\mvis$ distribution for signal with
$m_A$=140~GeV/$c^2$. Signal and background normalizations
correspond to the fit results for signal exclusion at 95\% CL.}
\label{fig:mVis}
\end{figure}

\begin{figure}
\includegraphics[width=0.95\columnwidth]{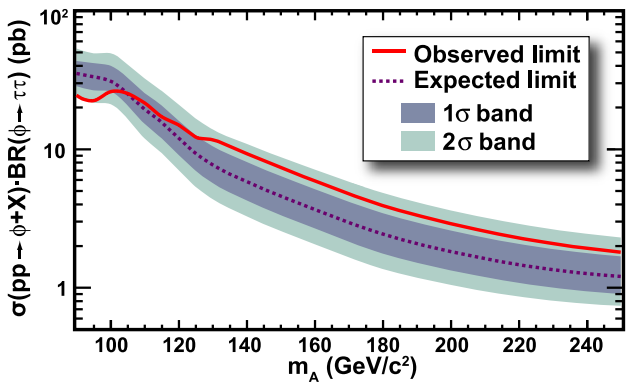}
\caption{Upper limits at 95\% CL on Higgs production cross section
times branching fraction to $\tau$ pairs. The expected limits from
the pseudo-experiments are also shown.} \label{fig:limits}
\end{figure}

We observe 487 events, in agreement with $N_{SM}$.
To probe for possible Higgs signal we
perform binned likelihood fits of the partially reconstructed mass
of the di-$\tau$ system ($\mvis$) defined as the invariant mass of
the visible tau decay products and $\met$. The backgrounds are
allowed to float within limits set by Gaussian constraints
corresponding to the systematic uncertainties in trigger
efficiencies, particle identification, production cross sections,
PDF's, event cuts, and luminosity measurement. Potential
differences in $\mvis$ shapes between data and the MC simulation
in different channels are treated as systematic uncertainties.
We create signal and background $m_{vis}$ templates with the MC
energy scales shifted from the nominal values according to the
uncertainties, and study the effect on hypothetical cross section
measurements. The deviations from the results obtained with the
nominal templates are parameterized in terms of Higgs mass and
input cross section.
An example fit for $m_A=140$~GeV/$c^2$ is shown in
Figure~\ref{fig:mVis}.
We observe no signal evidence for $m_A=90-250$~GeV/$c^2$, and set
exclusion limits at 95\% CL on $\sigma(p\bar{p}\to \phi + X) \times
BR(\phi \to \tau \tau)$ as shown in Figure~\ref{fig:limits}.
The sensitivity of the limit-setting procedure is determined
from MC simulations assuming no signal. The $m_{vis}$ shape
uncertainty leads to 15\% (5\%) deterioration of the limits for
the low (high) end of the considered $m_A$ region. The observed
limits range from 24.4~pb for $m_A=90$~GeV/$c^2$, to 9.3~pb for
$m_A=140$~GeV/$c^2$, to 1.8~pb for $m_A=250$~GeV/$c^2$.

\begin{figure}
\includegraphics[width=0.95\columnwidth]{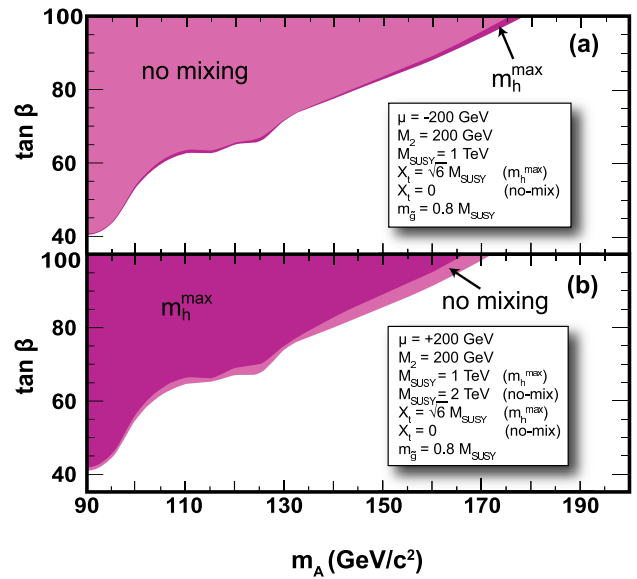}
\caption{Excluded regions in the $\tan\beta$ vs $m_A$ plane for
the $m_h^{max}$ and no-mixing scenarios with (a) $\mu<0$ and (b)
$\mu>0$.} \label{fig:tanBeta}
\end{figure}

Using the theoretical predictions for the MSSM Higgs production
and decay to $\tau$ pairs we interpret the limits
on $\sigma(p\bar{p}\to \phi + X)\times BR(\phi \to \tau\tau)$
as exclusions of parameter regions in the $\tan\beta$ {\sl vs} $m_A$
plane. The cross sections are obtained from SM calculations and
scaling factors $\sigma_{MSSM}/\sigma_{SM}$ accounting for the
modified  Higgs couplings~\cite{procedure}. The cross sections for
gluon fusion mediated by a $b$-quark loop are calculated with the
HIGLU program~\cite{higlu}. The corresponding values for
$b\bar{b}\to \phi + X$ are taken from~\cite{harlander}.
The scaling factors and $BR(\phi\to\tau\tau)$ are calculated
with the FeynHiggs program~\cite{feynhiggs}. They depend on $m_A$, $\tan\beta$,
the $SU(2)$ gaugino mass parameter $M_2$, the SUSY mass scale $M_{SUSY}$,
the squark mixing parameter $X_t$, the gluino mass $m_{\tilde{g}}$, and
the Higgs mixing parameter $\mu$. We consider four
benchmarks~\cite{benchmarks}: the $m_h^{max}$ and no-mixing
scenarios, with $\mu>0$ and $\mu<0$.
The excluded $\tan\beta$ vs $m_A$ regions are shown in
Figure~\ref{fig:tanBeta}.

The LEP experiments have excluded $m_A \lesssim 93$ GeV/$c^2$, and
higher-mass $A$ for small $\tan\beta$~\cite{LEP}.
Our search is complementary, providing sensitivity in the large
$\tan \beta$ region. The excluded parameter space in the
$\tan\beta$ vs $m_A$ plane for $\mu<0$
is similar to the D{\O} results obtained in the $\phi \to
b\bar{b}$ decay mode~\cite{d0}, and extends to higher $m_A$.
Moreover, our results in the $\phi\to \tau\tau$ channel
allow us to set comparable exclusions for scenarios with $\mu>0$,
as the lower production cross sections are compensated by an increase
in $BR(\phi \to \tau \tau)$.

We thank
A. Belyaev, M. Carena,  J. Gunion,
S. Heinemeyer,    W. Kilgore,
S. Mrenna, M. Spira, C. Wagner, G. Weiglein, and S. Willenbrock
for illuminating
discussions on the theory of MSSM Higgs production and decays.

We thank the Fermilab staff and the technical staffs of the
participating institutions for their vital contributions. This
work was supported by the U.S. Department of Energy and National
Science Foundation; the Italian Istituto Nazionale di Fisica
Nucleare; the Ministry of Education, Culture, Sports, Science and
Technology of Japan; the Natural Sciences and Engineering Research
Council of Canada; the National Science Council of the Republic of
China; the Swiss National Science Foundation; the A.P. Sloan
Foundation; the Bundesministerium f\"ur Bildung und Forschung,
Germany; the Korean Science and Engineering Foundation and the
Korean Research Foundation; the Particle Physics and Astronomy
Research Council and the Royal Society, UK; the Russian Foundation
for Basic Research; the Comisi\'on Interministerial de Ciencia y
Tecnolog\'{\i}a, Spain; in part by the European Community's Human
Potential Programme under contract HPRN-CT-2002-00292; and the
Academy of Finland.


\begin{thebibliography}{99}

\bibitem{higgs}
    P.W. Higgs, Phys. Lett. {\bf 12}, 132 (1964); Phys. Rev. Lett. {\bf 13}
    508 (1964); Phys. Rev. {\bf 145}, 1156 (1966).
\bibitem{martin} S.P. Martin, hep-ph/9709356 (1997), and references therein.

\bibitem{mssm} D.J.H. Chung {\sl et al.}, Phys.Rept. {\bf 407}, 1
(2005), and references therein.

\bibitem{dawson}
    S. Dawson, A. Djouadi, and M. Spira, Phys. Rev. Lett. {\bf 77}, 16 (1996).

\bibitem{maltoni}
        F. Maltoni, Z.Sullivan, and S. Willenbrock, Phys. Rev. D {\bf 67}, 093005 (2003).

\bibitem{harlander}
        R.V. Harlander and W.B. Kilgore, Phys. Rev. D {\bf 68}, 013001 (2003).

\bibitem{run1} D. Acosta {\sl et al.} (CDF Collaboration), hep-ex/0506042 (2005).

\bibitem{hmtt} D. Acosta {\sl et al.} (CDF Collaboration),  Phys. Rev. Lett. (to be published);
hep-ex/0506034 (2005).

\bibitem{CDFdetector}
    D. Acosta {\sl et al.} (CDF Collaboration), Phys. Rev. D {\bf 71}, 032001 (2005).

\bibitem{luminosity}
    S. Klimenko, J. Konigsberg, and T.M. Liss, FERMILAB-FN-0741 (2003).

\bibitem{tauTrig}
    A. Anastassov {\sl et al.}, Nucl. Instrum. Methods A {\bf 518}, 609 (2004).

\bibitem{pt} We define transverse
momentum ($p_T$) and transverse energy ($E_T$) as $p_T=p\sin\theta$
and $E_T=E\sin\theta$.

\bibitem{metDef} D. Acosta {\sl et al.} (CDF Collaboration), hep-ex/0504053 (2005).

\bibitem{pythia}
    T. Sj\"{o}strand {\sl et al.}, Comput. Phys. Commun. {\bf 135}, 238 (2001).
    We use PYTHIA v. 6.215.

\bibitem{cteq5} H.L. Lai {\sl et al.} (CTEQ Collaboration), Eur. Phys. J. C {\bf 12}, 375
(2000).

\bibitem{tauola}
    Z. Was {\sl et al.}, Nucl. Phys. Proc. Suppl. {\bf 98}, 96 (2001).

\bibitem{geant}
    R. Brun and F. Carminati, CERN Programming Library Long Writeup {\bf W5013} (1993).

\bibitem{zCrossSection}
    D. Acosta {\sl et al.} (CDF Collaboration), Phys. Rev. Lett. {\bf 94}, 091803 (2005).

\bibitem{pdfUncert} We compare the acceptances obtained with the CTEQ5 and CTEQ6 PDF sets.
                    In the latter case we also include the effect of the eigen-vector variations.

\bibitem{procedure} M. Carena, S. Heinemeyer, G. Weiglein, and C.E.M. Wagner,
FERMILAB-PUB-05-370-T (2005).


\bibitem{higlu}
    M. Spira, Nucl. Instrum. Methods A {\bf 389}, 357 (1997).

\bibitem{feynhiggs}
    S. Heinemeyer, W. Hollik, and G. Weiglein, Eur. Phys. J. C {\bf 9}, 343 (1999);
    Comput. Phys. Commun. {\bf 124}, 76 (2000).

\bibitem{benchmarks}
    M. Carena, S. Heinemeyer, C.E.M. Wagner, and G. Weiglein,
    hep-ph/9912223 (1999);
    Eur. Phys. J. C {\bf 26}, 601 (2003).

\bibitem{LEP} ALEPH, DELPHI, L3, and OPAL Collaborations, LHWG-Note 2004-01 (2004).

\bibitem{d0} V. Abazov {\sl et al.} (D{\O} Collaboration), hep-ex/0504018 (2005).

\end{thebibliography}
\end{document}